\documentclass[11pt,a4paper]{article}
\usepackage{jheppub}

\usepackage{multirow, graphicx,amssymb,url,mathrsfs,amsmath}
\usepackage{wrapfig,boxedminipage,setspace,subfigure,epsfig}
\usepackage{amsxtra,amstext,latexsym,dsfont,amsfonts}
\usepackage{color,eucal}
\usepackage[dvipsnames]{xcolor}
\usepackage{float}
\usepackage{slashed,comment}
\usepackage{kotex}
\usepackage{tensor}

\usepackage{indentfirst}

\newcommand{\abs}[1]{\left| #1 \right|}

\title{Weyl semimetal from non-inertial observers}

\author[a]{Wen-Bin Pan}
\author[b,c]{and Ya-Wen Sun}

\emailAdd{panwb@ihep.ac.cn}
\emailAdd{yawen.sun@ucas.ac.cn}

\affiliation[a]{Institute of High Energy Physics, Chinese Academy of Sciences,\\19B Yuquan Road, Shijingshan District, Beijing 100049, China}
\affiliation[b]{School of Physical Sciences, University of Chinese Academy of Sciences,\\ Zhongguancun East Road 80, Beijing 100190, China}
\affiliation[c]{Kavli Institute for Theoretical Sciences, University of Chinese Academy of Sciences,\\ Zhongguancun East road 80, Beijing 100190, China}

\abstract{We show that a reference frame transformation could turn a topologically trivial Dirac fermion into a topologically nontrivial Weyl semimetal. This is elucidated by the transformation of the Dirac equation into the equation for Weyl semimetals through specific infinitesimal local Lorentz transformations of the orthonormal basis. This kind of transformation, interpreted as a change of reference frame, could induce an observational effect that an axial gauge field and/or a vector U(1) gauge field appears effectively, which are in fact inertial forces in the non-inertial frame.
The precise local Lorentz transformations and the movement of observers needed to realize the two additional fields are provided respectively.
This novel effect can be viewed as a generalization of the effect found in relativistic hydrodynamics that topologically trivial modes in an inertial frame could become topologically nontrivial observed by a special non-inertial observer.
}


\arxivnumber{}

\begin{document}
\maketitle
\flushbottom


\section{Introduction}

In recent decades, topological states of matter have been studied extensively in condensed matter physics \cite{Wen:2016ddy, Witten:2015aoa}.
Representative topological states include anomalous Hall effect, topological insulators, and topological superconductors, which are gapped \cite{Haldane:1988zza, Hasan:2010xy, Qi:2010qag, Hasan:2010hm, Wehling:2014cla}, and various kinds of topological semimetals such as Weyl/nodal line/Dirac semimetals, which are gapless topological states of matter \cite{Yan:2016euz, Liu:2018djq, Landsteiner:2019kxb,fang2016topological}. Soon after the discovery of these quantum topological states of matter, many classical topological states have also been found in various systems, e.g. sound, optical systems, and hydrodynamic systems, etc. \cite{lu2014topological, ozawa2019topological, zhang2018topological, souslov2019topological, green2020topological, delplace2017topological}

Recently an interesting observational effect related to topological states of matter was found in \cite{Liu:2020abb, Liu:2020ksx}, which states that in relativistic hydrodynamic systems, topologically trivial modes in an inertial frame could become topologically nontrivial observed by a special accelerating observer. This is a new observational effect of accelerating observers in addition to the famous Unruh effect \cite{Unruh:1976db, Crispino:2007eb}. In the Unruh effect, whether a state is vacuum or thermal depends on observers, and now it has been shown that whether a system is topologically trivial or non-trivial is also a property that could depend on observers. 

This discovery in the hydrodynamic system has also been generalized to the case with one extra U(1) conserved current and more complicated structures of topologically nontrivial modes were found \cite{Pan:2020nhx}. 
Moreover, a non-inertial frame version of AdS/CFT correspondence was developed. 
It is shown that for the same non-inertial observer, the hydrodynamics modes calculated from the holographic system match exactly to the dual hydrodynamic results \cite{Pan:2020nhx}. The underlying mechanism of why topological trivial states could become topologically nontrivial was also explained in \cite{Pan:2020nhx}. Physically, this is easier to understand as in non-inertial frames, there will be inertial forces that could produce the interaction terms needed in topological states of matter. From the point of view of the transformation of states between different reference frames, the explanation is that the topologically nontrivial states now seen in a non-inertial reference frame come from the topologically nontrivial states in the complex momentum space of the original inertial observer, which could not be seen from the original inertial frame.

As this interesting property of topologically trivial modes becoming topologically non-trivial observed by a specific accelerating observer was found first in hydrodynamic systems, we may ask if this is only a property of hydrodynamic systems or if this is a more general property associated with accelerating observers, i.e. the change of non-inertial reference frames. A first attempt would be to study fermionic systems, which are the most studied topological systems in condensed matter physics. In this paper, as a first step, we start from gapless topological semimetal states, the Weyl semimetal. The aim is to see if a trivial massive Dirac fermion in an inertial frame would become a topologically nontrivial Weyl semimetal state observed by a special non-inertial observer.

From a more technical point of view, we would see that it is expected that the property that topologically trivial states could become topologically nontrivial observed by an accelerating observer should in principle also be found in other systems. The reason is as follows. The observation of topologically non-trivial modes by specific accelerating observers in relativistic hydrodynamic systems relies on the existence of non-zero components of the Christoffel connection in non-inertial frames, i.e. inertial forces. In hydrodynamics, the equation of motion of the system is the conservation equation and when we transform from an inertial frame to a non-inertial frame, the original conservation equation for the stress-energy tensor $\partial_\mu T^{\mu \nu}=0$ would become $\nabla_\mu T^{\mu \nu}=0$ with the metric also transformed accordingly.
From the perspective of a non-inertial observer who observes and records physical quantities using the coordinates of the new reference frame, the additional connection terms in the covariant conservation equation would be seen as inertial forces, i.e. external interactions that modify the equation of motion.
Consequently, the equation of motion has extra interaction terms as seen by the non-inertial observer and topologically non-trivial modes could appear due to these interactions.
One could also expect that the connection terms, coming from the covariant derivatives, would also modify the equation of motion of other systems when considered in non-inertial frames.
Hence it is reasonable to believe that the effect of topologically trivial modes becoming topologically non-trivial is a property associated with non-inertial observers while is not specifically associated with relativistic hydrodynamics.

To answer this question and find this property of non-inertial observers in other systems, in this paper, we start from a topologically trivial fermionic system whose dynamics are governed by the trivial massive Dirac equation and see if a Weyl semimetal equation of motion could be realized in a non-inertial frame due to the inertial forces. 
Note that in fermionic systems, there is another kind of connection called spinor affine connection which determines the parallel transport of spinors living at different spacetime positions and corresponds to a spinorial covariant derivative.\footnote{Relevant formalism is introduced in section \ref{2.1}.}
Due to redundant degrees of freedom in the definition of the tetrad, i.e. the set of four vectors comprising an orthonormal basis, we could define the coordinates and the tangential frames independently for the accelerating observer. 
Thus we determine the observer's direction utilizing the tetrad rather than relying on coordinates and take this difference of $e_\mu^a$ compared to the original inertial spacetime one as also part of the reference frame transformation. When the tangential bases are in different directions compared to the coordinate system, we can view this as a rotation or boost of the bases. This is an equivalent way to defining a rotating/boost reference frame transformation 
and in this way, the rotation/boost of bases becomes more clear. We call this a local Lorentz transformation, which refers to the rotation and boost of the bases of the observer, which might be different at each spacetime point. This is a non-inertial frame transformation, though the word Lorentz is in the name. The spin connection could become non-zero after performing a local Lorentz transformation that makes the orthonormal basis spacetime dependent.

We will show that it is possible that the extra terms in the spin connection would produce exactly the interaction terms in a Weyl semimetal compared to ordinary massive Dirac fermions.
More specifically, the topologically trivial massive Dirac fermions would be observed to have the same behavior as a topologically nontrivial Weyl semimetal in some specific reference frames.
The bases of the new frame are related to the original ones by infinitesimal local Lorentz transformations. 
 We will demonstrate in this work how the massive Dirac equation can become the Weyl semimetal equation \eqref{Weyl} by a transformation of the reference frame and show exactly what kind of reference frame transformation is needed to realize the interacting terms. 

The paper is organized as follows. In section \ref{2} we review some basic ingredients regarding the tetrad formalism and the massive Dirac equation in general spacetimes as well as equations for a Weyl semimetal in the flat spacetime, which are needed for our computations. 
In section \ref{3}, we discuss how to find a Weyl semimetal-like topologically nontrivial system by performing a change of reference frame.
In section \ref{4} we look at the modified Dirac equation after changing to other reference frames 
and show that two external fields, an axial gauge field and an imaginary electromagnetic gauge field, would appear from specific local Lorentz transformations.
In section \ref{5}, we conclude the paper with some discussions and open questions about the relationship with strained Weyl semimetal.

\section{Review of the Weyl semimetal from Dirac equation}\label{2}

The primary objective of this paper is to investigate whether the phenomenon observed in hydrodynamic systems, where topologically trivial states in an inertial frame may transition to topologically nontrivial states when observed by an accelerating observer, can be identified in other systems. As a first step, we focus on fermionic systems and start from a topologically trivial massive Dirac equation to see if it could become a topologically nontrivial system in certain non-inertial frames. Thus we will need to study the Dirac equation in a non-inertial frame, i.e. with a metric different from the Minkowski one, and check if the extra terms from the nonzero components of the spin connection will produce the exact inertial forces that we need for a Weyl semimetal. We will first review the Dirac equation in general spacetime and then review the Lagrangian for a Weyl semimetal in flat spacetime in this section.

\subsection{Review of Dirac equation in general spacetimes}\label{2.1}

 In this subsection, we review some basics of the Dirac equation in general spacetimes following \cite{Collas:2018jfx} and specify our notations. In flat spacetime, the massive Dirac equation is
\begin{align}\label{Dirac}
	    i \gamma^\mu \partial_\mu \psi - m \psi = 0.
\end{align}
In order to write down the Dirac equation in a curved spacetime (or a flat spacetime with a non-Cartesian coordinate system), we employ the tetrad formalism.
A tetrad is a set of four orthonormal basis vectors that can be defined at each point of spacetime.
By definition, the tetrad, $e^\mu_a$, obey
\begin{equation}\label{tetrad}
\begin{aligned}
      	g_{\mu \nu}e^\mu_a e^\nu_b = \eta_{ab}\\
        \eta_{ab}e^a_\mu e^b_\nu = g_{\mu \nu}. 
\end{aligned} 
\end{equation}
We use Greek indices $\mu, \nu$ to indicate spacetime coordinates and Latin indices ($a,b=0,1,2,3$) to specify the tangential space indices. 
{$\eta_{ab}=\text{diag}(-1,1,1,1)$} is the Minkowski spacetime metric in Cartesian coordinates. 
There are redundant degrees of freedom in the tetrad because a local Lorentz transformation (tetrad rotation)
\begin{align}
		\Tilde{e}^\mu_a=\Lambda^b_a(x) e^\mu_b
\end{align}
does not change \eqref{tetrad} for a fixed metric $g_{\mu \nu}$, where $\Lambda^b_a(x)$ is the local Lorentz transformation matrix and $\eta_{cd} \Lambda^c_a(x) \Lambda^d_b(x) = \eta_{ab}$.
We can view different choices for the tetrad as corresponding to different reference frames, i.e. observers with axes pointing to different directions. These correspond to the kind of rotating non-inertial reference frame transformation as the direction of the axes are position {and time} dependent. These different reference frames are related to each other by local Lorentz transformations. An equivalent way to deal with rotating non-inertial reference frame transformations here is to impose the condition that the bases of the tangential space have to be in the same direction as the coordinate space bases and here we choose to use the tangential space bases to denote the different rotating non-inertial frames as this makes the choice of the non-inertial observer more clear. 

We also need to introduce the spacetime dependent gamma matrices $\Bar{\gamma}^\mu(x)$ which are related to the constant tangential space gamma matrices $\gamma^a$ by the relation
\begin{align}\label{gammax}
		\Bar{\gamma}^\mu(x)=e^\mu_a(x)\gamma^a.
\end{align}
The anti-commutation relations for $\gamma^a$ and $\Bar{\gamma}^\mu(x)$ are
\begin{align}
		\{\gamma^a,\gamma^b\}=2\eta^{ab}\mathbf{1},\\
        \{\Bar{\gamma}^\mu(x),\Bar{\gamma}^\nu(x)\}=2 g^{\mu \nu}\mathbf{1},
\end{align}
where $\mathbf{1}$ denotes the 4 dimensional identity matrix.

The spinor $\psi$ transforms as a scalar under general coordinate transformations but transforms non-trivially under local Lorentz transformations as 
\begin{align}
	    \Tilde{\psi}=L(x) \psi,
\end{align}
 where $L(x)$ is the spinor representative of a local Lorentz transformation.
The derivative of a spinor $\psi$ transforms like a vector under general coordinate transformations but does not transform like a spinor under local Lorentz transformations (tetrad rotations) since 
\begin{align}
	    \partial_\mu \Tilde{\psi}=L(x) \partial_\mu \psi+\partial_\mu L(x) \psi.
\end{align}
We need to introduce a spinor affine connection $\Gamma_\mu$ to define the covariant derivative for a spinor as\footnote{To be precise, the connection is actually a matrix with two spinor indices, $(\Gamma_\mu)^b_a$.}
\begin{align}
	    D_\mu \psi= \mathbf{1} \partial_\mu \psi+ \Gamma_\mu \psi,
\end{align}
and require the covariant derivative of a spinor to transform like a spinor 
\begin{align}
	    \tilde{D}_{\mu} \tilde{\psi}=L D_{\mu} \psi.
\end{align}
This is the spinorial analog of the spacetime covariant derivative $\nabla$.
We further impose the requirement that the spinor affine connection is metric compatible, 
\begin{align}
	    D_{\mu} g^{\alpha \beta} \mathbf{1} =0.
\end{align}
Parallel to the case of the spacetime affine connection $\Gamma^\lambda_{\mu \nu}$, the form of spinor affine connection $\Gamma_\mu$ is fixed by the metric compatibility to be 
\begin{align}
	    \Gamma_\mu=-\frac{i}{4}\omega_{\mu ab}\sigma^{ab},
\end{align}
where $ \sigma^{ab}=\frac{i}{2}\left [\gamma^a, \gamma^b\right ]$ and $\omega_{\mu ab}=e_{a \lambda}\nabla_\mu e^\lambda_b$.

After obtaining $\Gamma_\mu$, we could write down the Dirac equation in general spacetime in terms of the spinorial covariant derivative as
\begin{align}\label{DiracC}
	    i \Bar{\gamma}^\mu D_\mu \psi - m \psi = 0
\end{align}
where $\Bar{\gamma}^\mu=e^\mu_a(x)\gamma^a$ and $D_\mu=\mathbf{1} \partial_\mu + \Gamma_\mu$.
\subsection{Review on Weyl semimetals in the flat spacetime}
Weyl semimetal is a very interesting form of gapless topological quantum matter\cite{Vafek:2013mpa, Hosur:2013kxa, Armitage:2017cjs}.
They exhibit many exotic features including the chiral anomaly, the appearance of surface states (Fermi-arcs), and exotic transport phenomena such as the anomalous Hall conductivity (see e.g. \cite{Hosur:2013kxa, Landsteiner:2015pdh, Landsteiner:2019kxb} and references therein).
They are characterized by crossing nodes (Weyl points) in the Brillouin zone at which the conduction and valence bands touch.
The Weyl points are robust under small perturbations indicating that they are topologically nontrivial.

\begin{figure}[htbp]
    \centering
    \includegraphics[scale=0.35]{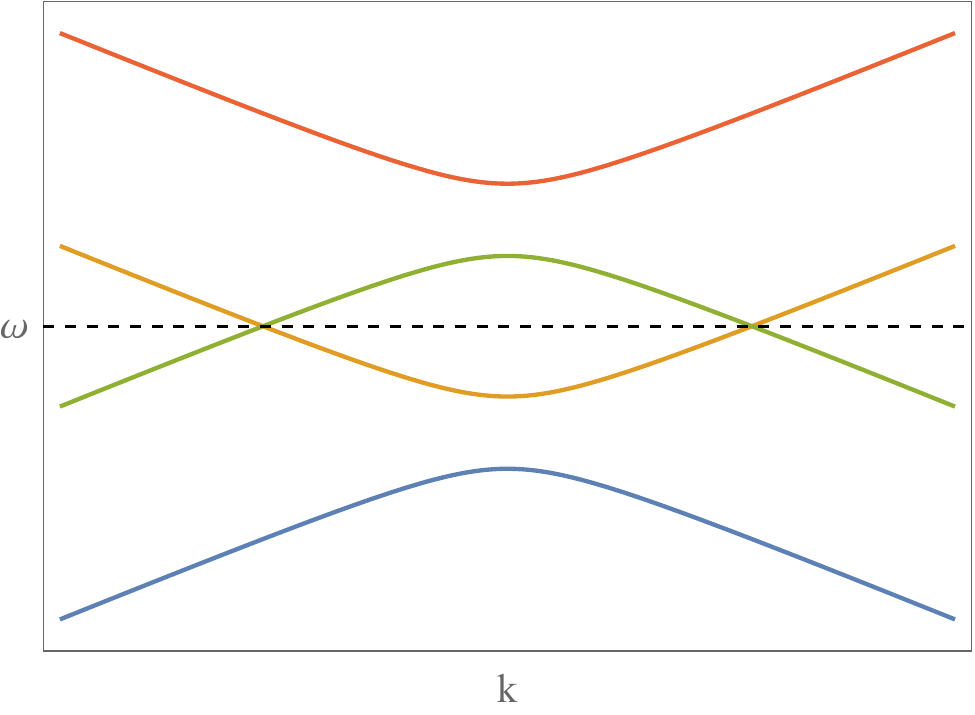}
    \caption{The energy spectrum for the Hamiltonian \eqref{WeylH} for $b_\mu=(0,0,0,b_3)$. A conduction band and a valence band cross at a pair of Weyl nodes which are separated by the axial gauge field.}
    \label{Weylspectrum}
\end{figure}

A field theoretical model that describes a Weyl semimetal takes the form of a
“Lorentz breaking” Dirac equation \cite{Grushin:2012mt,Grushin:2019uuu}
\begin{align}\label{Weyl}
	    (i \gamma^\mu (\partial_\mu +  i A_\mu) -b_\mu \gamma^5 \gamma^\mu - m )\psi = 0,
\end{align}
where $A_\mu$ is the electromagnetic gauge field, $b_\mu$ is the axial gauge field, $\gamma^\mu$ are the Gamma matrices, and $\gamma^5=i\gamma^0\gamma^1\gamma^2\gamma^3$. This system has the following real-time action \footnote{The gauge field $A_\mu$ is turned off for now.}  
\begin{align}
	S=\int d^4 x \; \Bar{\psi}(i \gamma^\mu \partial_\mu -b_\mu \gamma^5 \gamma^\mu - m )\psi.
\end{align}
After a Legendre transformation to the Lagrangian, the corresponding Hamiltonian is
\begin{align}\label{WeylH}
	H=\int d^3 x \; \psi^\dagger (-i \gamma^0 \gamma^j \partial_j +b_\mu \gamma^0 \gamma^5 \gamma^\mu + m \gamma^0)\psi.
\end{align}
The time component of the axial gauge field $b_0$ breaks the parity symmetry of this Hamiltonian while the spatial components $b_j (j=1,2,3)$ break the time-reversal symmetry.
In the following sections, we refer to the term $b_0 \gamma^5 \gamma^0$ as the parity-breaking term and $b_j \gamma^5 \gamma^j \, (j=1,2,3)$ as the time-reversal symmetry-breaking term.
To be concise, one could choose $b_\mu=(0,0,0,b_3)$\footnote{Note that the Weyl nodes exist only when the four-vector $b_\mu$ is spacelike otherwise the spectrum is gapped and describes an insulator.} and $\vec{k} =(0,0,k_3)$.
The spectrum for this Hamiltonian is then
\begin{equation}
	\omega = \pm (b_3 + s \sqrt{(k_3^{2}+m^{2})}),\\	
\end{equation}
where $s=\pm 1$. Figure \ref{Weylspectrum} shows this spectrum of Weyl semimetal.
Because the time-reversal symmetry is broken, the Kramers degeneracy is also broken, and we have four non-degenerate energy bands.
The two energy bands with $s=1$ are always gapped and the bands corresponding to $s=-1$ are gapped if $m>\abs{b_3}$. When $m<\abs{b_3}$, the bands with $s=-1$ have a pair of crossing points called Weyl nodes and effective Weyl fermion excitation can be obtained around these points.

\section{Weyl semimetal from non-inertial observers}
\label{3}

	In this section, we show how to find a Weyl semimetal-like topologically nontrivial system in a non-inertial frame from a trivial massive Dirac equation in the original inertial frame. 
As we have already explained, we will use $e^\mu_a$ to denote the change of the spacetime and the corresponding accelerating observer if it exists. We start from the massive Dirac equation in a general spacetime with an arbitrary tetrad $e^\mu_a$ and try to find which explicit tetrad could give extra terms that a Weyl semimetal action requires, compared to the inertial case. Then we will see if this required tetrad corresponds to the one of a non-inertial observer and if yes, what is this specific accelerating observer.

The spacetime for the inertial observer is flat 
 with the metric $\eta_{\mu\nu}$ and the orthonormal basis $e^\mu_a$ to be an identity matrix $\delta^\mu_a$ with row index $\mu$ and column index $a$. In a non-inertial reference frame, the metric would be different from $\eta_{\mu\nu}$ and we assume that $e^\mu_a$ becomes
	\begin{align}\label{evar}
	    e^\mu_a=\delta^\mu_a+ f^\mu_a,
	\end{align}
	where all the components in $f^\mu_a$ are small and may depend on the spacetime coordinates $x$. $e^a_\mu$ is the inverse of $e_a^\mu$ and we have 
	\begin{align}\label{eivar}
		e^a_\mu=\delta^a_\mu- f^a_\mu,
	\end{align} where we have ignored higher order terms of $f^\mu_a$ as we have assumed $f^\mu_a\ll 1$.
	From \eqref{tetrad}, the metric becomes
	\begin{align}
		g_{\mu \nu}&=(\delta^a_\mu- f^a_\mu)(\delta^b_\nu- f^b_\nu)\eta_{ab}\notag\\
		&=\eta_{\mu \nu}-(f_{\mu \nu}+f_{\nu \mu})+O(f^2).\label{gvar}
	\end{align}
	The variation of the metric is $h_{\mu \nu}=-(f_{\mu \nu}+f_{\nu \mu})$ to the first order of $f$. The Christoffel connection is
	\begin{align}\label{Con}
		\Gamma^{\lambda}_{\mu \nu}=\frac{1}{2}\eta^{\lambda \rho}(\partial_\mu h_{\rho \nu}+\partial_\nu h_{\rho \mu}-\partial_\rho h_{\mu \nu})+O(f^2),
	\end{align}
	which is of the first order in $f$.

	Next, we calculate the extra terms of the Dirac equation in the metric background \eqref{evar}  compared to the inertial frame one with the metric being $\eta_{\mu\nu}$ in terms of the matrix $f$. We assume that \eqref{evar} corresponds to the spacetime of a non-inertial observer so the terms associated with $f$ are inertial forces.
	Put \eqref{evar}\eqref{eivar}\eqref{gvar}\eqref{Con} into the Dirac equation in general spacetime \eqref{DiracC} and Fourier transform the  spinor in the new reference frame $\psi$ to the corresponding momentum space $\psi(x) = \int d^4 k \, \Tilde{\psi}(k) e^{i k_\mu x^\mu}$, we get 
 \begin{equation}
     \begin{aligned}
     i\gamma^a e^\nu_a \partial_\nu \psi(x) 
     &= \int d^4 k \, i\gamma^a e^\nu_a(ik_\nu \tilde{\psi}(k)e^{ik_\mu x^\mu})\\
     &=-\int d^4 k \, \gamma^a k_a \tilde{\psi}(k)e^{ik_a x^a}\\
     &\to i\gamma^a \partial_a \psi(x).
	\end{aligned}
 \end{equation}
 Note that we have defined $x^a: = e_\mu^a x^\mu$, $k_a: = e^\mu_a k_\mu$ and $\partial_a := e^\mu_a \partial_\mu$. The components of $k_a$ can be thought of as the projection of four-vector $k$ onto the tetrad. Since the tetrad defines our reference frame, $k_a$ (instead of $k_\mu$) is the observable in the new reference frame.
	Then we denote the additional term originating from the spin connection to be $\Sigma= \frac{1}{4}\gamma^a e^\nu_a \omega_{\nu bc} \sigma^{bc}$.
	After expanding $\Sigma$ to the first order in $f$, we obtain
	\begin{align}\label{Sigma}
	    \Sigma=\frac{1}{4}(\partial_a f_{[cb]}+\partial_{[b}f_{,a,c]}+\partial_{[b}f_{c]a})\gamma^a \sigma^{bc},
	\end{align}
	where 
 \begin{equation}
     \begin{aligned}
	    \partial_a f_{[cb]}=\frac{1}{2}(\partial_a f_{cb}-\partial_a f_{bc}),\\
	    \partial_{[b}f_{,a,c]}=\frac{1}{2}(\partial_b f_{ac}-\partial_c f_{ab}),\\
	    \partial_{[b}f_{c]a}=\frac{1}{2}(\partial_b f_{ca}-\partial_c f_{ba}).\\
	\end{aligned}
 \end{equation}
	Thus eqn.\eqref{Dirac} becomes 
	\begin{align}
	    i \gamma^a \partial_a \tilde{\psi} - m \tilde{\psi} +\Sigma \tilde{\psi}= 0
	\end{align}
	after changing the tangential space basis vectors.
	This $\Sigma$ term is the first order deviation from the original inertial frame Dirac equation in another frame with $e^\mu_a$ in \eqref{evar} and this term corresponds to the inertial force. Note that for reference frame transformations, we should have non-inertial forces in a non-inertial frame as long as it is not an inertial frame transformation of a global Lorentz transformation, different from a coordinate transformation, though their transformation relations appear similar.

\subsection{The corresponding non-inertial observer}\label{3.1}

An immediate question that comes after this result is what kind of transformation is needed to realize the extra term $\Sigma$ in the equation of motion. Note that we have explained that here we choose the reference frame to be determined by the tetrad, thus as will be shown explicitly later, a coordinate transformation $x^\mu \to x'^\mu(x)$ as used in \cite{Liu:2020ksx} in hydrodynamics would not be related to a reference frame transformation here, while the local Lorentz transformation works as we require all observable observed by the observer are those projected to the tangential basis. This transformation, besides transforming the spacetime coordinates, changes the orthonormal basis of the observer and could be thought of as a local ``rotation" of the basis vectors.
The coordinate transformation is a change in the spacetime index $\mu$ in $e^\mu_a$, while the local Lorentz transformation amounts to a mixing of the basis vectors which are labeled by $a$ in our setup
	\begin{align}
	    e^a_\mu \to \Lambda^a_b(x) e^b_\mu,
	\end{align}
where $\Lambda^a_b(x)$ is a Lorentz transformation matrix and it is ``local" because it is dependent on the spacetime position $x$. 
With a coordinate transformation $x^\mu \to x'^\mu(x)$ and a local Lorentz transformation $\Lambda^a_b(x)$, the most general transformation to the inertial reference frame $\delta_a^\mu$ is
\begin{align}\label{cor-lor}
	    e^\mu_a = \frac{\partial x'^\mu}{\partial x^\nu} \Lambda^a_b \delta^\nu_a.
\end{align}
The infinitesimal form of this combined transformation is\footnote{Quadratic and higher-order terms in \eqref{infi} are omitted}
\begin{align}\label{infi}
	    \delta^\mu_a \to e^\mu_a = (\delta^\mu_\nu + \partial_\nu \xi^\mu)(\delta^b_a+\lambda^b_a)\delta^\nu_b=\delta^\mu_a+\partial_\nu \xi^\mu \delta^\nu_b+\lambda^b_a \delta^\nu_b
\end{align}
where $\xi^\mu$ denote an infinitesimal coordinate displacement $x'^\mu \to x^\mu+\epsilon^\mu$, $\Lambda^b_a=\delta^b_a+\lambda^b_a+O(\lambda)^2$ and $\lambda^b_a$ obeys $\lambda_{ab}+\lambda_{ba}=0$\footnote{The tangential space indices should be lowered (or raised) by $\eta_{ab}$ ($\eta^{ab}$).}

However, here we show that a coordinate transformation does not contribute to $\Sigma$ at all, due to the reason we mentioned above: all observables are projected to the $a$ basis. In this way $\Sigma= \frac{1}{4}\gamma^a e^\nu_a \omega_{\nu bc} \sigma^{bc}$ is a scalar that does not change under any coordinate transformation.
One could also check this by calculating $\Sigma$ from \eqref{Sigma} directly.
Under an infinitesimal coordinate transformation $x'^\mu \to x^\mu+\epsilon^\mu$, the tetrad changes according to 
\begin{align}
	    \delta^\nu_a \to e^\mu_a = \frac{\partial x'^\mu}{\partial x^\nu} \delta^\nu_a = (\delta^\mu_\nu + \partial_\nu \xi^\mu)\delta^\nu_a,
\end{align}
and the corresponding $f_{ab}$ is 
\begin{align}\label{fab}
    f_{ab}=(\partial_\nu \xi_\mu) \delta^\nu_a \delta^\mu_b:= \partial_a \xi_b.
\end{align}
After putting \eqref{fab} into \eqref{Sigma}, we find that $\Sigma$ vanishes due to the fact that the partial derivatives commute with each other.

 
On the contrary, the Local Lorentz transformation does contribute to $\Sigma$ and change the spectrum.
This is because the $\Sigma$ term comes from the spin affine connection and, as introduced in section \ref{2.1}, has the same status as the Christoffel connection in a coordinate transformation.
Under the local Lorentz transformation \eqref{lambda}, although the metric keeps invariant to be $\eta_{\mu \nu}$, the tetrad is not homogeneous (i.e. spacetime position dependent). This leads to a non-zero spin affine connection and thus a non-zero $\Sigma$.
Thus the local Lorentz transformation is what we need in a Dirac fermion system. Physically this is easy to understand as in our calculation, we have used the tetrad to indicate our reference frame, while the coordinates are not really associated with reference frames, different from the case of the hydrodynamic topological modes.

In the next, We first show the results for the local Lorentz transformation and leave the derivation of the results to section \ref{4}.
We find three kinds of infinitesimal transformation that would give the extra term required in the equation of motion of a Weyl semimetal.
They are: 

a) A boost in the $x$ direction\footnote{To see that it is really an infinitesimal boost, we recall that a Lorentz transformation matrix representing a boost in $x$ direction with rapidity $\phi$ is $\Lambda=\begin{pmatrix}
\cosh{\phi}&  \sinh{\phi}\\
\sinh{\phi}&  \cosh{\phi}\\
\end{pmatrix}$. With $\phi \ll 1$, we have $\Lambda \approx \begin{pmatrix}
1&  \phi\\
\phi&  1\\
\end{pmatrix}$. Similar approximations are applied in (b) and (c).} 
\begin{align}\label{(a)}
\Lambda^a_b=
\begin{pmatrix}
1&  -\frac{by}{2}&  0& 0\\
-\frac{by}{2}&  1&  0& 0\\
0&  0&  1& 0\\
0&  0&  0& 1
\end{pmatrix};
\end{align}

b) A boost in the $y$ direction
\begin{align}\label{(b)}
\Lambda^a_b=
\begin{pmatrix}
1&  0&  -\frac{bx}{2}& 0\\
0&  1&  0& 0\\
-\frac{bx}{2}&  0&  1& 0\\
0&  0&  0& 1
\end{pmatrix};
\end{align}

c) A rotation around the $z$ direction
\begin{align}\label{(c)}
\Lambda^a_b=
\begin{pmatrix}
1&  0&  0& 0\\
0&  1&  \frac{bt}{2}& 0\\
0&  -\frac{bt}{2}&  1& 0\\
0&  0&  0& 1
\end{pmatrix};
\end{align}
respectively, where $\Lambda^a_b=\delta^a_b+f^a_b$. Here we have imposed the condition that $b t$, $b x$, and $b y$ are small so that higher-order terms could be ignored. This would require that we are studying the system in a relatively finite volume. In this limit, these matrices can be approximately viewed as three types of infinitesimal local Lorentz transformations respectively: a boost in the $x$ direction with rapidity $-\frac{by}{2}$, a boost in the $y$ direction with rapidity $-\frac{bx}{2}$, and a rotation around the $z$ direction with angular velocity $\frac{b}{2}$.
Thus they are really infinitesimal Lorentz transformations on the orthonormal basis in the tangential space, although the transformations on different spacetime points are different, which implies that these are not inertial frame transformations like ordinary Lorentz transformations but non-inertial reference frame transformations.
We will show what the new reference frame looks like and describe the motion of the corresponding observer in section \ref{3.2}.


Substituting any of the choices of a), b), or c) into \eqref{Sigma} gives
	\begin{align}\label{SigmaW}
	\Sigma=
	\begin{pmatrix}
	0&  0&  b/2& 0\\
	0&  0&  0& -b/2\\
	b/2&  0&  0& 0\\
	0&  -b/2&  0& 0
	\end{pmatrix}.
	\end{align}
This $\Sigma$ is identical to $\frac{b}{2} \gamma^5 \gamma^3$.	
Therefore, the time-reversal symmetry-breaking term required in the equation of motion for the Weyl semi-metal here appears in our assumed non-inertial reference frame as an inertial force
	\begin{align}\label{A5z}
	i \gamma^a \partial_a \tilde{\psi} - m \tilde{\psi} + \frac{b}{2} \gamma^5 \gamma^3 \tilde{\psi}= 0.
	\end{align}
The Hamiltonian in this frame is then 
 \begin{align}
	H=\int d^3 x \; \psi^\dagger (-i \gamma^0 \gamma^a \partial_a -\frac{b}{2} \gamma^0 \gamma^5 \gamma^3 + m \gamma^0)\psi
\end{align}
and the corresponding spectrum is
 \begin{equation}
	\omega = \pm \sqrt {\frac{b^{2}}{4}+ k_1^{2}+ k_2^{2}+ k_3^{2}+ m^{2}+ s\sqrt{b^{2}(k_3^{2}+m^{2})}}.\\
 \end{equation}
Therefore, one needs only a local tangential space Lorentz transformation like \eqref{(a)}, \eqref{(b)} or \eqref{(c)} to change the spectrum of the Dirac equation \eqref{Dirac} to the spectrum of Weyl semi-metal \eqref{A5z}.
Physically, this means if an observer rotates and boosts their prescribed $4$ orthogonal directions precisely as \eqref{lambda}, a fermion system that was originally described by the Dirac equation in flat spacetime would be seen as a Weyl semi-metal. 

Note that we have assumed that $\abs{f^{\mu}_a}\ll 1$ and this condition here requires $bt\ll 1$, i.e. $b\ll 1/t$. This means that we have to observe the system within a finite interval of time and $b$ has to be small compared to $1/t$. At the same time, from the spectrum above, to get a Weyl semimetal state with two Weyl nodes, we need to have $b>2m$, which means if we observe the system in a finite but long enough time interval, then $b$ must be small enough, and for us to have Weyl nodes, we need $m$ to be small too, i.e. $m\ll 1/t$.

Note that though a coordinate transformation does not change the reference frame in our framework, it could change the look of $f$ and $\Lambda$, though the reference frame transformation might be the same when $f$ and $\Lambda$ look different. Consequently, the results \eqref{(a)}, \eqref{(b)} and \eqref{(c)} are not all the matrices that would generate the extra $\Sigma$ term of Weyl semimetal.
Other possible choices are the combination of these three and coordinate transformations which do not contribute to $\Sigma$.
For example, one could start with a simple guess of $f^\mu_a$
\begin{equation}\label{f}
       f^{\mu}_{a}=
	\begin{pmatrix}
	0&  0&  0& 0\\
	0&  0&  bt& 0\\
	0&  0&  0& 0\\
	0&  0&  0& 0
	\end{pmatrix} 
\end{equation}
This choice could generate the same $\Sigma$ term as \eqref{SigmaW} and therefore realize the spectrum of Weyl semimetal.
The new guess \eqref{f} does not originate from a coordinate transformation since there is no solution for $\xi^\mu$ in the equation 
\begin{align}
    \partial_\mu \xi^\nu \delta_a^\mu=f_a^\nu.
\end{align}
Obviously, it is also not induced by any local Lorentz transformation which preserves the metric.
Therefore, in order to change the reference frame (tetrad) from $\delta^\mu_a$ to $e^\mu_a = \delta^\mu_a + f^\mu_a$, 
both transformations are needed as in \eqref{infi}.
We have just shown that the coordinate transformation does not contribute to the change of the Dirac equation.
But since the choice \eqref{f} does indeed generate an additional $\Sigma$ term, the contribution must come totally from the local Lorentz part of the transformation.

In the following, we decompose \eqref{f} into an infinitesimal coordinate transformation and an infinitesimal local Lorentz transformation, and thus extract the effective part from the full transformation.
We first deduce the form of the coordinate transformation from the variation of the metric.
Since the local Lorentz transformation does not affect the metric components $\eta_{\mu \nu}$ in coordinate space, the change of the metric comes totally from a coordinate transformation $\xi$:
	\begin{align}
	    \eta_{\mu \nu} \to \eta_{\mu \nu}-\partial_\mu \xi_\nu-\partial_\nu \xi_\mu.
	\end{align}
Compare this to the variation caused by metric variation $f^\mu_a$, i.e. eqn.\eqref{gvar}, we obtain
	\begin{align}\label{lie}
	    \partial_\mu \xi_\nu+\partial_\nu \xi_\mu = f_{\mu \nu}+f_{\nu \mu}.
	\end{align}
Solving $\xi_\mu$ from this equation with $f^\mu_a$ set to be \eqref{f}, we find the most general solution is 
\begin{equation}
    \begin{aligned}
	    \xi^0&=\frac{1}{2}bxy-c_1 x-c_2 y-c_3 z+i_0\\
	    \xi^1&=\frac{1}{2}bty-c_1 t+d_1 y-d_2 z+i_1\\
	    \xi^2&=\frac{1}{2}btx-c_2t-d_1 x+d_3 z+i_2\\
	    \xi^3&=-c_3t+d_2 x-d_3 y+i_3
	\end{aligned}
\end{equation}
In addition to $b$, there are ten possible small parameters in the solution. $c_1$, $c_2$, $c_3$ are parameters for three infinitesimal boosts, $d_1$, $d_2$, $d_3$ are parameters for three infinitesimal rotations and $i_0$, $i_1$, $i_2$, $i_3$ are parameters for four infinitesimal translations.
These together constitute an infinitesimal Poincaré transformation in coordinate space\footnote{Do not confuse it with the local Lorentz transformation of the tetrad.}.
Since the additional Poincaré transformation does not change the metric, we can choose to turn it off, which means we let all the ten parameters be zero. Thus we find the simplest solution to be
    \begin{align}\label{xi}
	    \xi^\mu=(\frac{b}{2}xy,\frac{b}{2}ty,\frac{b}{2}tx,0).
	\end{align}
The full solution is the simplest solution above in addition to an arbitrary infinitesimal Poincaré transformation.
	
Although the change in metric induced by \eqref{xi} matches with that caused by \eqref{f}, the change in $e^\mu_a$ is not fully determined by the coordinate transformation. 
The degrees of freedom of local Lorentz transformation in the tangential space must be taken into consideration. 
According to \eqref{cor-lor}, the additional local Lorentz transformation is
\begin{align}\label{lambda}
\Lambda^a_b(x)  = e^\mu_b \frac{\partial x^\nu}{\partial x'^\mu} \delta^a_\nu = (\delta^\mu_b+f^\mu_b)(\delta^\nu_\mu - \partial_\mu \xi^\nu)\delta^a_\nu=
\begin{pmatrix}
            1&  -\frac{by}{2}&  -\frac{bx}{2}& 0\\
-\frac{by}{2}&              1&   \frac{bt}{2}& 0\\
-\frac{bx}{2}&  -\frac{bt}{2}&              1& 0\\
            0&              0&              0& 1
\end{pmatrix}.
\end{align}
This is precisely a combination of the three types of infinitesimal local Lorentz transformations of our results \eqref{(a)}, \eqref{(b)} and \eqref{(c)}.


\subsection{Movement of the observer }\label{3.2}

In this section, we discuss the movement of the observer that could possibly observe a Weyl semimetal state. The motion and the corresponding tetrad chosen by the observer are described by the local Lorentz transformation.
As mentioned in the previous subsection, it is a combination of the following three kinds of motion and these three kinds of motion are shown in figure \ref{b3picture} :

\begin{figure}[htbp]\label{b3picture}
    \centering
    \subfigure[]{
    \begin{minipage}[b]{.3\linewidth}
    \centering
    \includegraphics[scale=0.22]{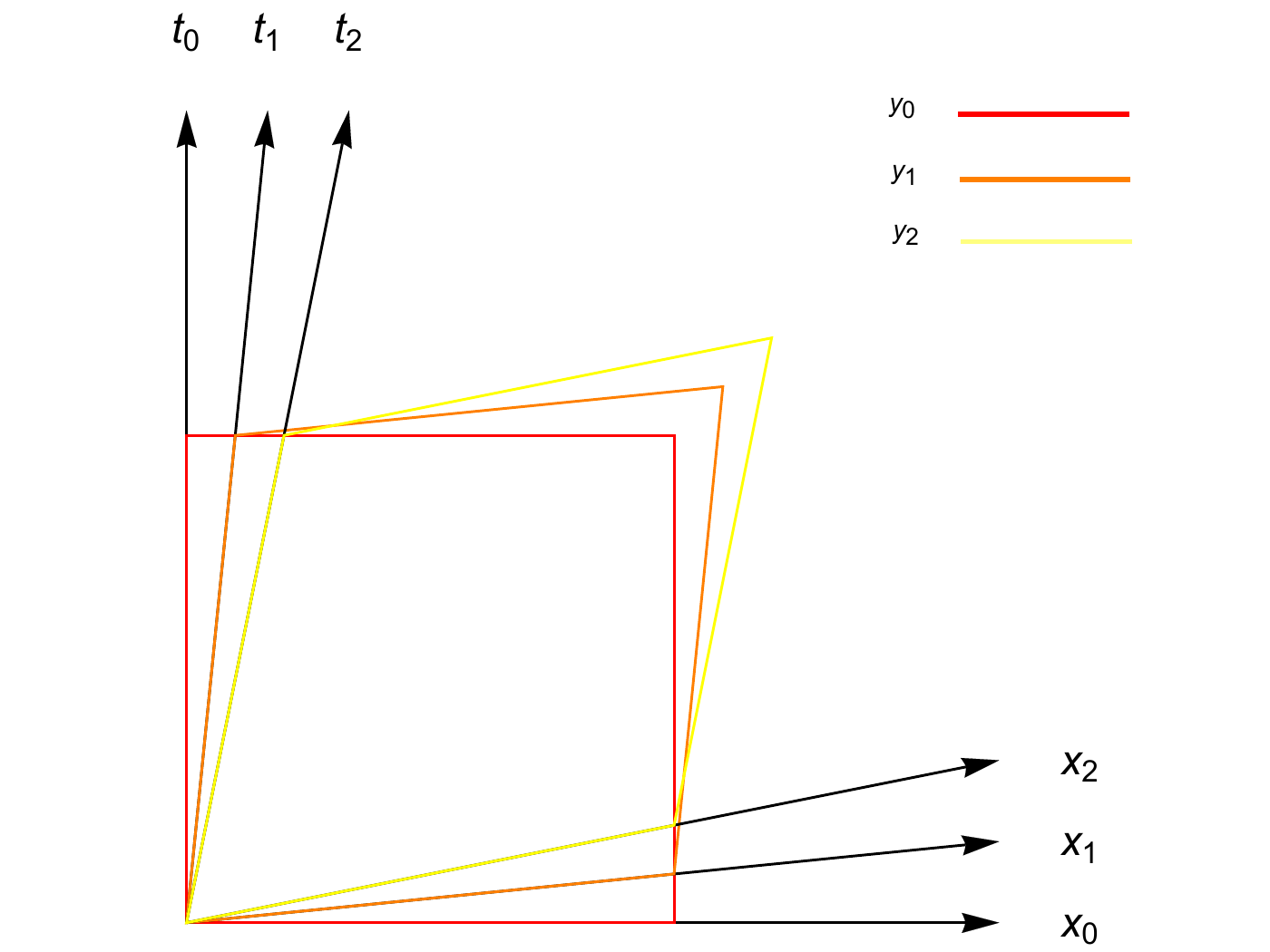}
    \end{minipage}
    }
    \subfigure[]{
    \begin{minipage}[b]{.3\linewidth}
    \centering
    \includegraphics[scale=0.22]{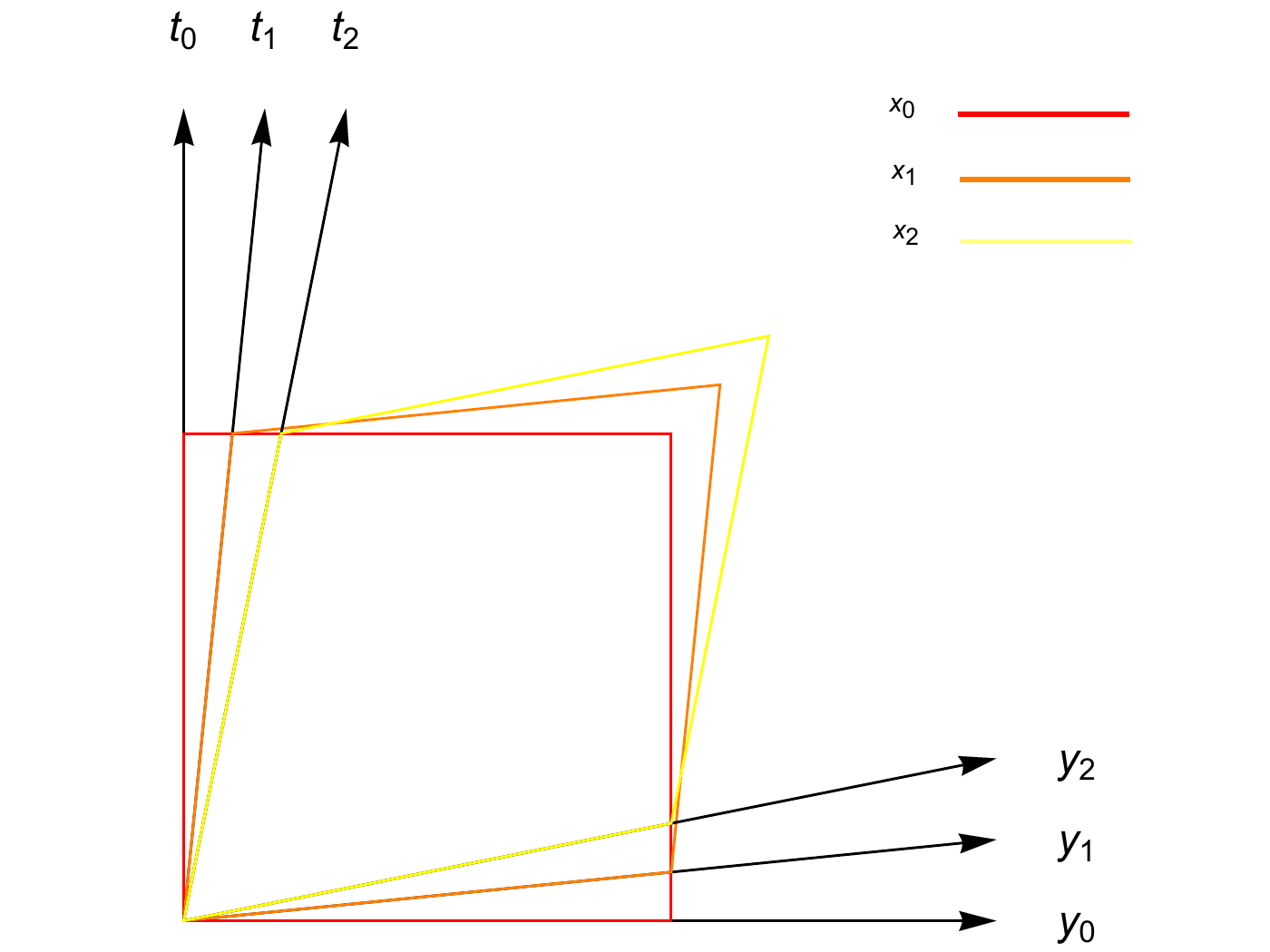}
    \end{minipage}
    }
    \subfigure[]{
    \begin{minipage}[b]{.3\linewidth}
    \centering
    \includegraphics[scale=0.22]{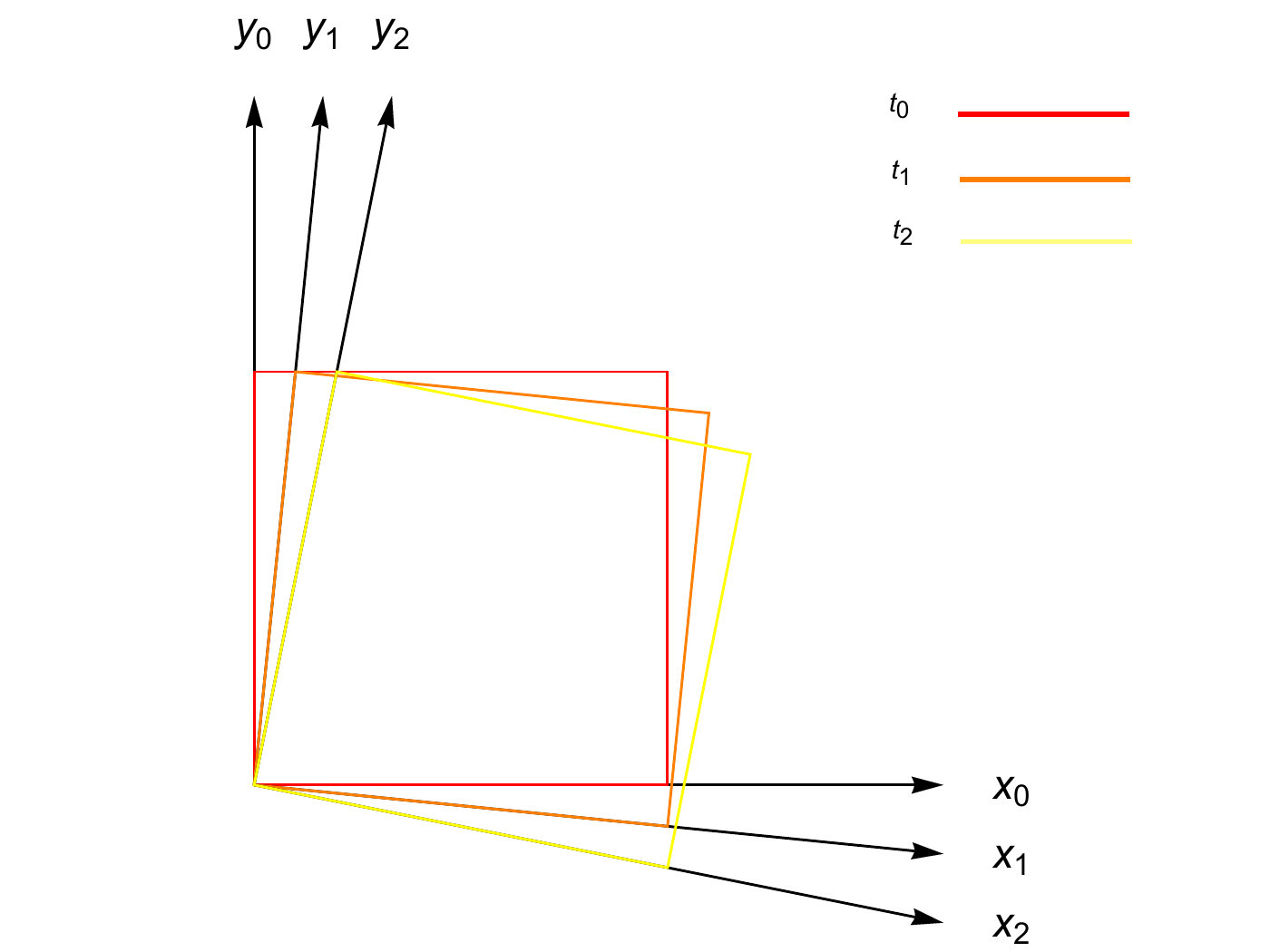}
    \end{minipage}
    }
    \caption{Three kinds of motion for observers that could possibly observe a Weyl semimetal state ($b=-0.1$). (a) Movement along the $x$ direction with velocity $-\tanh(\frac{by}{2})$. Different colors indicate three observers sitting at three different positions with $y_0$, $y_1$, and $y_2$ ($y_2>y_1>y_0>0$) who move with different velocities ($v_2>v_1>v_0=0$); (b) Movement along the $y$ direction with velocity $-\tanh(\frac{bx}{2})$. Different colors indicate three observers sitting at three different positions with $x_0$, $x_1$, and $x_2$ ($x_2>x_1>x_0>0$) who move with different velocities ($v_2>v_1>v_0=0$); (c) Rotation around the $z$ direction with constant angular velocity $\frac{b}{2}$. Different colors represent the position of a square observer at three different moments of time ($t_2>t_1>t_0>0$). 
    }    
\end{figure}

(a) A boost in the $x$ direction:
\begin{align}
\Lambda^a_b=
\begin{pmatrix}
1&  -\frac{by}{2}&  0& 0\\
-\frac{by}{2}&  1&  0& 0\\
0&  0&  1& 0\\
0&  0&  0& 1
\end{pmatrix}.
\end{align}
The observer needs to move along the $x$ direction with velocity $-\tanh(\frac{by}{2})$.
Moreover, as shown in figure \ref{b3picture} (a), the farther he/she is away in the $y$ direction the larger his/her velocity is. 

(b) A boost in the $y$ direction:
\begin{align}
\Lambda^a_b=
\begin{pmatrix}
1&  0&  -\frac{bx}{2}& 0\\
0&  1&  0& 0\\
-\frac{bx}{2}&  0&  1& 0\\
0&  0&  0& 1
\end{pmatrix}.
\end{align}
Motion (b) is the same as (a) except that the roles of $x$ and $y$ are exchanged. The observer moves along the $y$ direction with velocity $-\tanh(\frac{bx}{2})$, and as shown in figure \ref{b3picture} (b), the farther he/she is away in the $x$ direction the larger his/her velocity is. 

(c) Rotation around the $z$ direction:
\begin{align}
\Lambda^a_b=
\begin{pmatrix}
1&  0&  0& 0\\
0&  1&  \frac{bt}{2}& 0\\
0&  -\frac{bt}{2}&  1& 0\\
0&  0&  0& 1
\end{pmatrix}.
\end{align}
The observer rotates themselves around the $z$ direction with constant angular velocity $\frac{b}{2}$ as shown in figure \ref{b3picture} (c). In this case the observer could be a rigid object while in the first two cases the observer cannot be a rigid object. In the first two cases, the observer must either be elastic or consist of local observers positioned at various spatial points with different velocities.


\section{Axial chemical potential and the gauge potential term from non-inertial observers}\label{4}

In the previous section, the time-reversal symmetry-breaking term $-b \gamma^5 \gamma^3$ is realized by a special local Lorentz transformation, producing a Weyl semimetal state in a certain non-inertial reference frame.  
Actually, it is also possible to find more additional terms including the parity symmetry breaking term $-b_0 \gamma^5 \gamma^0$ from reference frame changes. 
Note the spectrum associated with a Dirac Hamiltonian adding only this term does not have Weyl nodes and is topologically trivial.
Let us study the extra term $\Sigma$ systematically here. 

Firstly, we assume the infinitesimal variation $f^\mu_a$ is linear in spacetime coordinates, i.e. it has the form of
\begin{align}\label{assumption}
f^\mu_a=\Tilde{f}^{\;\mu}_{a\;\nu}x^\nu
\end{align}
where $\Tilde{f}^{\;\mu}_{a\;\nu}$ is a first-order small quantity.
Note that one uses $\delta^b_\mu$ to switch coordinate indices of $\Tilde{f}^{\;\mu}_{a\;\nu}$  into basis vector indices and use $\eta_{ab}$($\eta^{ab}$) to lower(raise) basis vector indices.
The resulting first order $\Sigma$ is 
\begin{align}
\Sigma &=\frac{1}{4}( \Tilde{f}_{[cb]a}+\Tilde{f}_{a[cb]}+\Tilde{f}_{[c,a,b]})\gamma^a \sigma^{bc}\notag\\
       &=\frac{i}{4}( \Tilde{f}_{[cb]a}+\Tilde{f}_{a[cb]}+\Tilde{f}_{[c,a,b]})\gamma^a \gamma^b \gamma^c.
\end{align}
$\Sigma$ could be rewritten as a sum of three gamma matrices and one gamma matrix by using the anticommutation relations of the gamma matrices\footnote{ $\Tilde{f}_{[abc]} $is the antisymmetrization of $\Tilde{f}_{abc}$, $\Tilde{f}_{[abc]}=\Tilde{f}_{abc}+\Tilde{f}_{bca}+\Tilde{f}_{cab}-\Tilde{f}_{bac}-\Tilde{f}_{acb}-\Tilde{f}_{cba}$. Note the normalization coefficient $3!$ is not included in our definition of antisymmetrization.}
\begin{align}\label{SigmaD}
\Sigma =-\frac{i}{4}\Tilde{f}_{[abc]}\gamma^a \gamma^b \gamma^c-\frac{i}{2}\eta^{bc}(\Tilde{f}_{abc}-\Tilde{f}_{bca})\gamma^a.
\end{align}
Note that we have decomposed $\Sigma$ into a set of linearly independent basis of gamma matrices.

Secondly, in order for $\Tilde{f}$ to describe a local Lorentz transformation so that the spacetime is still flat after the change in the tetrad, there is a constraint for $\Tilde{f}$:
\begin{align}\label{fconstrain}
\eta_{ab}(\delta^a_c+f^a_c)(\delta^b_d+f^b_d)=\eta_{cd}
\end{align}
which gives us $f_{ab}+f_{ba}=0$ or equivalently $\Tilde{f}_{abc}+\Tilde{f}_{bac}=0$.
This means $\Tilde{f}$ must be anti-symmetric with respect to the first two indices.
Taking this condition into account, \eqref{SigmaD} can be simplified to be
\begin{align}\label{SigmaD2}
\Sigma =-\frac{i}{2}\Tilde{f}_{\overline{abc}}\gamma^a \gamma^b \gamma^c-\frac{i}{2}\eta^{bc}\Tilde{f}_{abc}\gamma^a,
\end{align}
where $\overline{abc}$ represents a cyclic operation over the indices, $\Tilde{f}_{\overline{abc}}=\Tilde{f}_{abc}+\Tilde{f}_{bca}+\Tilde{f}_{cab}$.

At last, we find the most general term that could be realized from the assumption of \eqref{assumption} is determined by $8$ constants: $\Tilde{f}_{\overline{123}}$, $\Tilde{f}_{\overline{023}}$, $\Tilde{f}_{\overline{013}}$, $\Tilde{f}_{\overline{012}}$, $\eta^{bc}\Tilde{f}_{0bc}$, $\eta^{bc}\Tilde{f}_{1bc}$, $\eta^{bc}\Tilde{f}_{2bc}$, $\eta^{bc}\Tilde{f}_{3bc}$.

By matching this expression of $\Sigma$ with the parity symmetry-breaking term, $-b_0 \gamma^5 \gamma^0= i b_0 \gamma^1 \gamma^2 \gamma^3$, $b_0$ is related with $\Tilde{f}_{abc}$ by 
\begin{align}
b_0=-\frac{1}{2}\Tilde{f}_{\overline{123}}=-\frac{1}{2}(\Tilde{f}_{123}+\Tilde{f}_{231}+\Tilde{f}_{312}).
\end{align}
$\Tilde{f}_{abc}$ that fulfills this equation would create the parity symmetry-breaking term in the equation of motion.
All the possibilities for $\Tilde{f}_{abc}$ are generated by three simple local tangential space Lorentz transformations:

i) $-\Tilde{f}_{123}=\Tilde{f}_{213}=2b_0$,
\begin{align}
f^\mu_a=
\begin{pmatrix}
0&  0&  0& 0\\
0&  0&  2b_0 z& 0\\
0&  -2b_0 z&  0& 0\\
0&  0&  0& 0
\end{pmatrix};
\end{align}

ii) $-\Tilde{f}_{231}=\Tilde{f}_{321}=2b_0$, 
\begin{align}
f^\mu_a=
\begin{pmatrix}
0&  0&  0& 0\\
0&  0&  0& 0\\
0&  0&  0& 2b_0 x\\
0&  0&  -2b_0 x& 0
\end{pmatrix};
\end{align}

iii) $-\Tilde{f}_{312}=\Tilde{f}_{132}=2b_0$,
\begin{align}
f^\mu_a=
\begin{pmatrix}
0&  0&  0& 0\\
0&  0&  0& 2b_0 y\\
0&  0&  0& 0\\
0&  -2b_0 y&  0& 0
\end{pmatrix}.
\end{align}
We have thus found that an axial chemical potential $b_0$ would appear effectively by a change of reference frames, even though it does not exist in the original homogenous frame.
They are three different rotations around the $z$, $x$, and $y$ axis with angular velocity $2b_0 z$, $2b_0 x$, and $2b_0 y$ respectively.


It is also easy to find the time reversal symmetry-breaking term $-b_j \gamma^5 \gamma^j (j=1,2,3)$ from \eqref{SigmaD2}, where
\begin{align}
b_1=-\frac{1}{2}\Tilde{f}_{\overline{023}}, \quad b_2=\frac{1}{2}\Tilde{f}_{\overline{013}}, \quad b_3=-\frac{1}{2}\Tilde{f}_{\overline{012}}.
\end{align}
As an example, we focus on the case of non-zero $b_3$, i.e. we turn on the term $\Tilde{f}_{\overline{012}}$ and set other $7$ constants to be zero.
It is clear that there are again three independent local Lorentz transformations that generate all the possibilities for non-zero $b_3$. They are i) $-\Tilde{f}_{012}=\Tilde{f}_{102}=2b_3$, a boost in the $x$ direction\footnote{Note that $f^\mu_a=\delta^\mu_d \eta^{bd} \Tilde{f}_{abc}x^c$.}; ii) $-\Tilde{f}_{201}=\Tilde{f}_{021}=2b_3$, a boost in the $y$ direction; and iii) $-\Tilde{f}_{120}=\Tilde{f}_{210}=2b_3$, a rotation around the $z$ direction.
A combination of these three matches precisely with \eqref{lambda}.
Thus we rediscover the result found in \ref{3.1}.

One could also find in \eqref{SigmaD2} the term with a single gamma matrix:  
\begin{align}
A_a \gamma^a=\frac{i}{2}\eta^{bc}\Tilde{f}_{abc}\gamma^a.
\end{align}
The coefficient $A_a$ may be interpreted as an electromagnetic gauge potential, although it is purely imaginary.
Thus by turning on $\eta^{bc}\Tilde{f}_{abc}$, a pure imaginary electromagnetic gauge potential effectively appears.
The energy spectrum becomes complex if there is a pure imaginary gauge field. This is consistent with our speculation in \cite{Pan:2020nhx} that a reference frame transformation may turn the frequency $\omega$ from real to complex. 

\section{Conclusion}\label{5}	

In this paper, we have studied the Dirac equation with perturbed tetrad in flat spacetime and showed that the Dirac equation can be deformed to the equation of Weyl semimetals by specific perturbations.
We have interpreted the infinitesimal local Lorentz transformations that perturb the original homogeneous tetrad to an inhomogeneous one as changes of reference frame for an observer.
Different reference frame changes, i.e. infinitesimal boosts and rotations of the tetrad, lead to different additional terms in the Dirac equation.
The most general infinitesimal local Lorentz transformations (under the assumption \eqref{assumption}) would create two additional effective fields: an axial gauge field 
\begin{align}
b_a=-\frac{1}{2}(\Tilde{f}_{\overline{123}}, \Tilde{f}_{\overline{023}}, -\Tilde{f}_{\overline{013}}, \Tilde{f}_{\overline{012}}),
\end{align}
and an imaginary U(1) gauge field
\begin{align}
A_a =\frac{i}{2}\eta^{bc}\Tilde{f}_{abc}.
\end{align}

This means the trivial Dirac fermions in a homogeneous (inertial) reference frame could be observed as topologically non-trivial Weyl semimetals if the observer chose an inhomogeneous (non-inertial) reference frame.
This brings up a new interesting effect that is analogous to what was found in relativistic hydrodynamic systems, i.e. topologically trivial modes could become topologically nontrivial observed by a special non-inertial observer.
Moreover, all we need to observe the non-trivial spectra are specific local Lorentz transformations, which could be tested in experiments in principle.
We expect that it is possible to observe the chiral anomaly and other exotic features of Weyl semimetal in those specific reference frames.

The effect we found has some similarity with the induction of elastic gauge fields in strained Weyl semimetal \cite{Cortijo:2015hlt, Cortijo:2016wnf, Kamboj:2019vgc, Chernodub:2019lhw, Bugaiko:2020jwt}. 
Around the two nodal points in strained Weyl semimetal, elastic lattice deformations couple to electronic excitations in the form of a gauge vector field \cite{Cortijo:2015hlt}.
Our change of reference frames may be interpreted as effective lattice deformations. An observer in the new frame would find that the lattice is no longer regular. 
Conversely, for any small elastic lattice deformation, there exists a reference change that no deformation exists in the new frame.
Thus we conjecture that the axial gauge field $b_\mu$ induced by an inhomogeneous reference frame may be also considered as an elastic gauge field.
However, the gauge field induced by elastic lattice deformations only exists effectively in the Weyl semimetal, while the gauge field induced by a change of reference frame turns a massive Dirac fermion into a Weyl semimetal state. 
It would be an interesting topic to study the deeper relation and quantitative matching between our findings and the strained Weyl semimetal in the future.

\section*{Acknowledgement}

We thank Matteo Baggioli, Hyun-Sik Jeong, Xuan-Ting Ji, Yuan-Chun Jing, Xing-Xiang Ju, Teng-Zhou Lai, Bo-Hao Liu, Yan Liu, and Yuan-Tai Wang for their helpful advice. 
This work was supported by the National Key R\&D Program of China (Grant No. 2018FYA0305800), the National Natural Science Foundation of China (Grant No. 12035016 and 12275275), the Strategic Priority Research Program of Chinese Academy of Sciences (Grant No. XDB28000000) and the Beijing Natural Science Foundation (Grant No. 1222031).

%


\bibliography{reference}
\bibliographystyle{JHEP}

\end{document}